\documentclass{article}

\usepackage{PRIMEarxiv}

\usepackage[utf8]{inputenc} 
\usepackage[T1]{fontenc}    
\usepackage{hyperref}       
\usepackage{url}            
\usepackage{booktabs}       
\usepackage{amsfonts}       
\usepackage{nicefrac}       
\usepackage{microtype}      
\usepackage{lipsum}
\usepackage{fancyhdr}       
\usepackage{graphicx}       
\usepackage{float}

\fancypagestyle{firstpage}{
  \fancyhf{} 
  \fancyfoot[C]{\small This work has been submitted to the IEEE for possible publication. Copyright may be transferred without notice, after which this version may no longer be accessible.}

}
  
\title{Runnable Directories: The Solution to the Monorepo vs. Multi-repo Debate}

\author{
  Shayan Ghasemnezhad, Samarth KaPatel, Sofia Nikiforova, \\[2pt] 
  \textbf{Giacinto Paolo (GP) Saggese, Paul Smith, Heanh Sok} \\[3pt]
  Causify AI \\[2pt]
  \texttt{\{shayan, s.kapatel, s.nikiforova, gp, paul, h.sok\}@causify.ai} \\
}

\begin{document}
\maketitle
\thispagestyle{firstpage}

\begin{abstract}
Modern software systems increasingly strain traditional codebase organization strategies. Monorepos offer consistency but often suffer from scalability issues and tooling complexity, while multi-repos provide modularity at the cost of coordination and dependency management challenges. As an answer to this trade-off, we present the Causify Dev system, a hybrid approach that integrates key benefits of both. Its central concept is the runnable directory --- a self-contained, independently executable unit with its own development, testing, and deployment lifecycles. Backed by a unified thin environment, shared helper utilities, and containerized Docker-based workflows, runnable directories enable consistent setups, isolated dependencies, and efficient CI/CD processes. The Causify Dev approach provides a practical middle ground between monorepo and multi-repo strategies, improving reliability and maintainability for growing, complex codebases.
\end{abstract}


\section{Introduction}
\label{sec:introduction}

Software development workflows are becoming more complex as they adapt to the demands of large-scale systems and modern collaborative development practices. As teams and codebases grow, companies face the challenge of organizing both effectively. When it comes to structuring the codebase, two main approaches emerge: monorepos and multi-repos\cite{castile2023}. Monorepos consolidate all code into a single repository, simplifying version control but carrying a risk of scalability and maintainability issues. Conversely, multi-repos store the code in logically separated repositories, easier to manage and deploy but more difficult to keep in sync.

Here, we present \textbf{Causify Dev system}, an alternative hybrid solution: a modular system architecture built around \emph{runnable directories}. Although independent, these directories maintain cohesion through shared tooling and environments, offering a straightforward and scalable way to organize the codebase while ensuring reliability in development, testing, and deployment.

\section{Current landscape}
\label{sec:current_landscape}

\subsection{Monorepo approach}
\label{sec:monorepo_approach}

The monorepo approach involves storing all code for multiple applications within a single repository. This strategy has been popularized by large tech companies like Google\cite{potvin2016}, Meta\cite{kadluczka2024}, Microsoft\cite{visual2018} and Uber\cite{zeino2017}, proving that even codebases with billions of lines of code can be effectively managed in a single repository. The key benefits of this approach include:

\begin{itemize}
\item Consistency in environment: with everything housed in one repository, there's no risk of projects becoming incompatible due to conflicting versions of third-party packages.
\item Simplified version control: there is a single commit history, which makes it easy to track and, if needed, revert changes globally.
\item Reduced coordination overhead: developers work within the same repository, with easy access to all code, shared knowledge, tools and consistent coding standards.
\end{itemize}

However, as monorepo setups scale, users often face significant challenges. A major downside is long continuous integration and continuous delivery (CI/CD) build times, as even small changes can trigger massive rebuilds and tests throughout the entire codebase. To cope with this, extra tooling, such as \href{https://buck2.build/}{Buck} or \href{https://bazel.build/}{Bazel}, must be configured, adding complexity to workflows. Even something as simple as searching and browsing the code becomes more difficult, often requiring specialized tools and plug-ins.

Additionally, when everything is located in one place, it is harder to separate concerns and maintain clear boundaries between projects. Managing permissions also becomes more difficult when only selected developers should have access to specific parts of the codebase.

\subsection{Multi-repo approach}
\label{sec:multirepo_approach}

The multi-repo approach involves splitting code across several repositories, with each one dedicated to a specific module or service. This modularity allows teams to work independently on different parts of a system, making it easier to manage changes and releases for individual components. Each repository can evolve at its own pace, and developers can focus on smaller, more manageable codebases.

However, the multi-repo strategy comes with its own set of challenges, particularly when it comes to managing dependencies and ensuring version compatibility across repositories. For instance, different repositories might rely on two different versions of a third-party package, or even conflicting packages, making synchronization complex or, in some cases, nearly impossible. In general, propagating changes from one repository to another requires careful coordination. Tools like \href{https://www.jenkins.io/}{Jenkins} and \href{https://github.com/features/actions}{GitHub Actions} help streamline CI/CD pipelines, but they often struggle when dealing with heterogeneous environments.

\subsection{What is needed}
\label{sec:what_is_needed}

An ideal strategy would combine the best of both worlds:

\begin{itemize}
\item The modularity of multi-repos, to keep the codebase scalable and simplify day-to-day development processes.
\item The environment consistency of monorepos, to avoid synchronization issues and prevent errors that arise from executing code in misaligned environments.
\end{itemize}

Both are achieved through our proposed hybrid approach, described in detail below.

\section{Causify Dev approach}
\label{sec:causify_dev_approach}

\subsection{Runnable directory}
\label{sec:runnable_directory}

The core concept of our approach is a \textbf{runnable directory} --- a self-contained, independently executable directory with code, equipped with a dedicated DevOps setup. A GitHub repository is thus a special case of a runnable directory. Developers typically work within a single runnable directory for a given application, enabling them to test and deploy code without affecting other parts of the codebase.

\begin{figure}
\centerline{\includegraphics[scale=0.6]{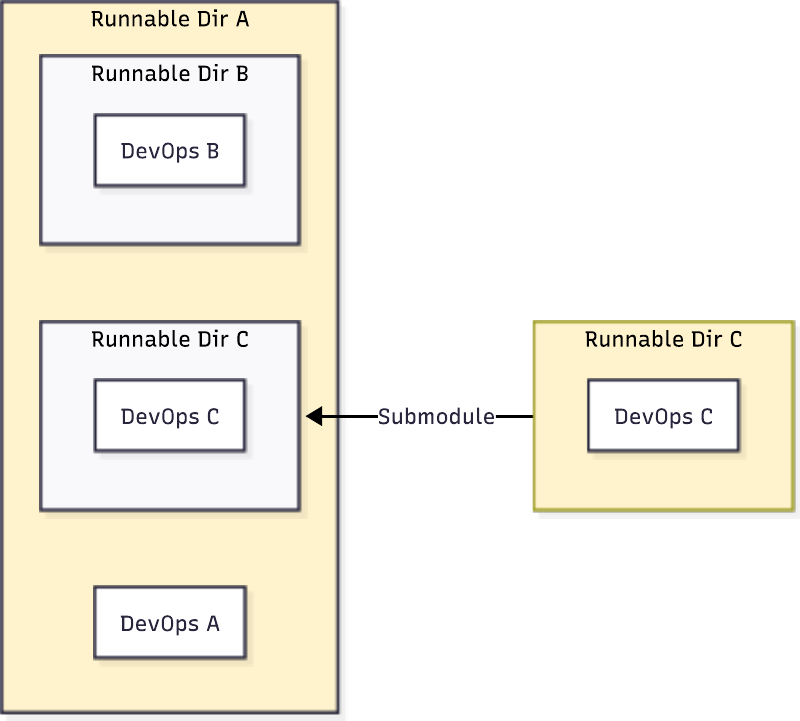}}
\caption{Sample architecture of runnable directories.}
\label{fig:rundir}
\end{figure}

A runnable directory can contain other runnable directories as subdirectories. For example, Figure~\ref{fig:rundir} depicts three runnable directories: \textit{A}, \textit{B}, and \textit{C}. Here, \textit{A} and \textit{C} are repositories, with \textit{C} incorporated into \textit{A} as a submodule, while \textit{B} is a subdirectory within \textit{A}. This setup provides the same accessibility as if all the code were hosted in a single monorepo. Note that each of \textit{A}, \textit{B}, and \textit{C} has its own DevOps pipeline, which is a key feature of our approach.

Each runnable directory contains a YAML file that defines its configuration parameters, such as identifiers, Docker container details, and Amazon S3 storage information. These parameters are readily accessible by the code within the directory, if needed.

\subsection{Docker}
\label{sec:docker}

\subsubsection{Container-driven environment}
\label{sec:container_driven_environment}

Docker is the backbone of our containerized development environment. Every runnable directory contains Dockerfiles that allow it to build and run its own Docker containers, which include the code, its dependencies, and the runtime system. All of our Docker containers are versioned, with their version history stored in changelog files.

This Docker-based approach addresses two important challenges. First, it ensures consistency by isolating the application from variations in the host operating system or underlying infrastructure. Our containers are compatible with different operating systems (Linux, macOS, Windows Subsystem for Linux) and support multi-architecture builds (e.g., x86 and ARM). Second, a specific package (or package version) can be added to the container of a particular runnable directory without affecting other parts of the codebase. This prevents ``bloating'' the environment with packages required by all applications --- a common issue in monorepos --- while also effectively mitigating the risk of conflicting dependencies, which can arise in a multi-repo setup.

We simplify and standardize all workflows associated with containers --- including building, testing, retrieving, and deploying --- by introducing Makefile-like tools based on the Python \textit{invoke} package.

\subsubsection{Stages of container development}
\label{sec:stages_of_container_development}

Our approach supports multiple stages for container release:

\begin{itemize}
\item Local: used to work on updates to the container; accessible only to the developer who built it.
\item Development: used by all team members in day-to-day development of new features.
\item Production: used to run the system by end users.
\end{itemize}

This multi-stage workflow enables seamless progression from testing to system deployment.

\subsubsection{Container interaction}
\label{sec:container_interaction}

\begin{figure}
\centerline{\includegraphics[scale=0.6]{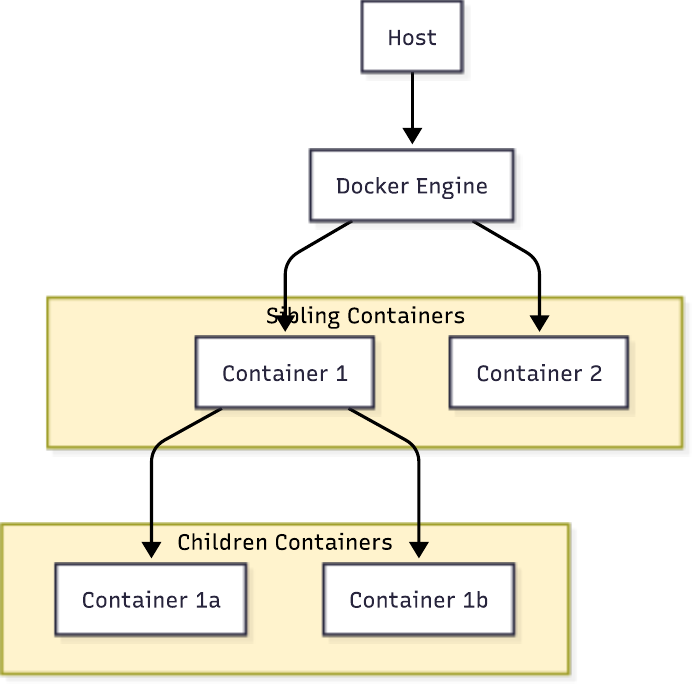}}
\caption{Docker container flow.}
\label{fig:containers}
\end{figure}

Our systems ensure smooth interaction between different containers in our infrastructure. Thus, development, testing, and deployment of multi-container applications are supported through Docker Compose. It is also possible to run a container within another container's environment in a Docker-in-Docker setup. In this case, children containers are started directly inside a parent container, allowing nested workflows or builds. Alternatively, sibling containers can run side by side and share resources such as the host's Docker daemon, enabling inter-container communication and orchestration. Both scenarios are illustrated by Figure~\ref{fig:containers}.

\subsection{Thin environment}
\label{sec:thin_environment}

\begin{figure}[H]
\centerline{\includegraphics[scale=0.6]{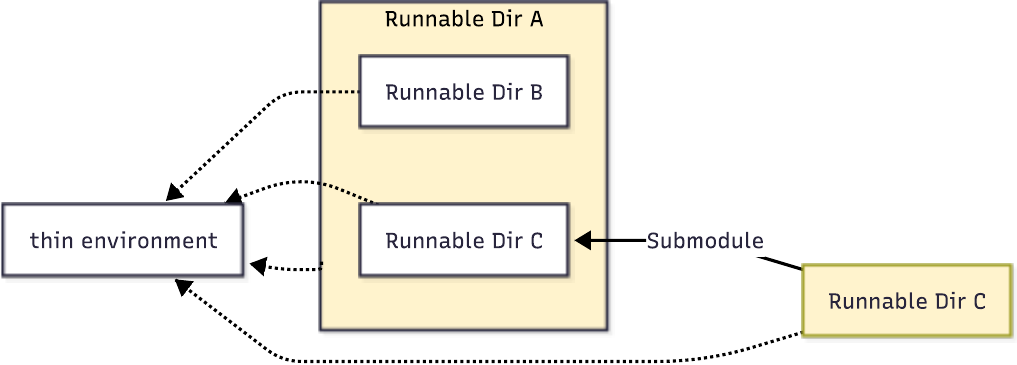}}
\caption{Thin environment shared across multiple runnable directories.}
\label{fig:thinenv}
\end{figure}

To bootstrap development workflows, we use a thin client that installs a minimal set of essential dependencies, such as Docker and invoke, in a lightweight virtual environment. A single thin environment is shared across all runnable directories, which minimizes setup overhead (see Figure~\ref{fig:thinenv}). This environment contains everything that is needed to start development containers, which are in turn specific to each runnable directory. With this approach, we ensure that development and deployment remain consistent across different systems (e.g., server, personal laptop, CI/CD).

\subsection{Submodule of ``helpers''}
\label{sec:submodule_of_helpers}

\begin{figure}
\centerline{\includegraphics[scale=0.6]{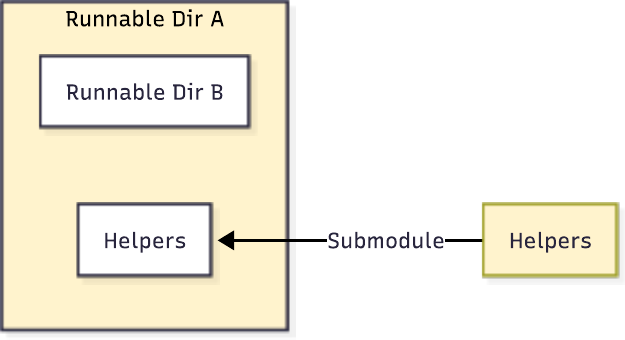}}
\caption{``Helpers'' submodule integrated into a repository.}
\label{fig:helpers}
\end{figure}

All Causify repositories include a dedicated ``helpers'' repository as a submodule (see Figure~\ref{fig:helpers}). This repository contains common utilities and development toolchains, such as the thin environment, Linter, Docker, and invoke workflows. By centralizing these resources, we eliminate code duplication and ensure that all teams, regardless of the project, use the same tools and procedures.

Additionally, it hosts symbolic link targets for files that must technically reside in each repository but are identical across all of them (e.g., license and certain configuration files). Manually keeping them in sync can be difficult and error-prone over time. In our approach, these files are stored exclusively in ``helpers'', and all other repositories utilize read-only symbolic links pointing to them. This way, we avoid file duplication and reduce the risk of introducing accidental discrepancies.

The ``helpers'' submodule also includes a set of Git hooks used to enforce policies across our development process, including Git workflow rules, coding standards, security and compliance, and other quality checks. These hooks are installed by default when the user activates the thin environment. They perform essential checks such as verifying the branch, author information, file size limits, forbidden words, Python file compilation, and potential secret leaks.

\subsection{Executing tests}
\label{sec:executing_tests}

\begin{figure}[H]
\centerline{\includegraphics[scale=0.6]{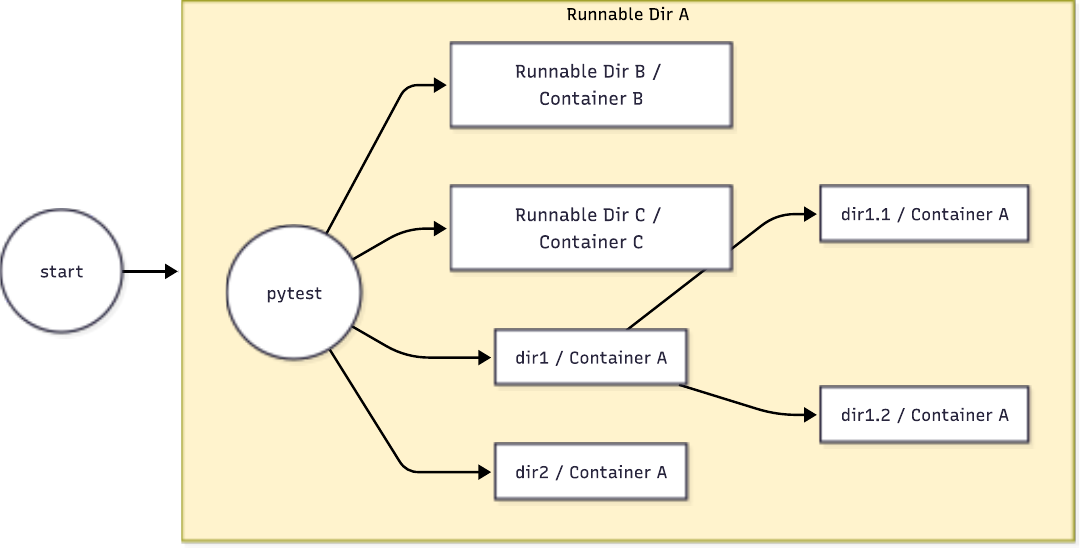}}
\caption{Recursive test execution in dedicated containers.}
\label{fig:tests}
\end{figure}

Our system features robust testing workflows that leverage the containerized environment for comprehensive code validation. End-to-end and unit tests are executed by \textit{pytest} inside Docker containers to ensure consistency across development and production environments, preventing discrepancies caused by variations in host system configurations. Both Docker-in-Docker and sibling container setup are supported. In the case of nested runnable directories, tests are executed recursively within each directory's corresponding container, which is automatically identified (see Figure~\ref{fig:tests}). As a result, the entire test suite can be run with a single command, while still allowing tests in subdirectories to use dependencies that may not be compatible with the parent directory's environment.

\subsection{Dockerized executables}
\label{sec:dockerized_executables}

Sometimes, installing a toolchain inside a development container is not justified, especially when it is large or used only occasionally. In such cases, we use \emph{dockerized executables} where the tool and its dependencies are packaged into a slim Docker image, and the command is executed inside a short-lived container. We launch this container with the repository mounted, run the tool, and then remove the container once the task is complete; the versioned image stays cached for reuse, avoiding repeated setup.

This approach prevents the development environment from being bloated with rarely used dependencies while preserving reproducibility and version control of the toolchain. When needed (e.g., during test execution), a dockerized executable can also run inside another Docker container using either the children or sibling container approach.

\section{Discussion}
\label{sec:discussion}

Causify's approach presents a strong alternative to existing code organization solutions, offering scalability and efficiency for both small and large systems.

The proposed modular architecture is centered around runnable directories, which operate as independent units with their own build and release lifecycles. This design bypasses the bottlenecks common in large monorepos, where centralized workflows can slow down CI/CD processes unless specialized tools like Buck or Bazel are used. By leveraging Docker containers, we ensure consistent application behavior across development, testing, and production environments, avoiding problems caused by system configuration discrepancies. Dependencies are isolated within each directory's dedicated container, reducing the risks of issues that tight coupling or package incompatibility might create in a monorepo or a multi-repo setup.

Unlike multi-repos, runnable directories can utilize shared utilities from ``helper'' submodules, eliminating code duplication and promoting consistent workflows across projects. They can even reside under a unified repository structure which simplifies codebase management and reduces the overhead of maintaining multiple repositories. With support for recursive test execution spanning all components, runnable directories allow for end-to-end validation of the whole codebase through a single command, removing the need for testing each repository separately.

There are, however, several challenges that might arise in the adoption of our approach. Teams that are unfamiliar with containerized environments may need time and training to effectively transition to the new workflows. The reliance on Docker may introduce additional resource demands, particularly when running multiple containers concurrently on development machines. This would require further optimization, possibly aided by customized tooling. These adjustments, while ultimately beneficial, can add complexity to the system's rollout and necessitate ongoing maintenance to ensure seamless integration with existing CI/CD pipelines and development practices.

\section{Future directions}
\label{sec:future_directions}

Looking ahead, there are several areas where the proposed approach can be improved. One direction is the implementation of dependency-aware caching to ensure that only the necessary components are rebuilt or retested when changes are made. This would reduce the time spent on development tasks, making the overall process more efficient. Further optimization could involve designing our CI/CD pipelines to execute builds, tests, and deployments for multiple runnable directories in parallel, which would allow us to take full advantage of available compute resources.

Another opportunity lies in addressing occasional Docker image size issues. When a runnable directory depends on code located in its parent, production builds may end up copying the entire upstream directory structure, resulting in unnecessarily large images. A potential solution is to automatically identify the specific files a runnable directory requires and include only those in the final image by generating an appropriate .dockerignore configuration. Such tooling would eliminate the need for manual pruning, which is currently time-consuming and error-prone.

Additional measures can also be taken to enhance security. Integrating automated container image scanning and validation before deployment would help guarantee compliance with organizational policies and prevent vulnerabilities from entering production environments. In addition, fine-grained access controls could be introduced for runnable directories in order to safeguard sensitive parts of the codebase. These steps will bolster both the security and efficiency of our workflows as the projects continue to scale.

\bibliographystyle{unsrt}  
\bibliography{references}  

\end{document}